\documentclass{preprint}

\title{Automatic Identification of Compounds in Molecular Mixtures from Liquid-Phase Infrared Spectra}

\author{
Yannah J.U. Melle,$^{1,2}$
Thanh Nguyen,$^{3}$
Jeffrey Lopez,$^{3,*}$
Daniel Schwalbe-Koda$^{1,*}$\\
\vspace{1em}
\normalfont{
\small
$^{1}$Department of Materials Science and Engineering, University of California, Los Angeles, CA, USA \\
$^{2}$Department of Chemistry and Biochemistry, University of California, Los Angeles, CA, USA \\
$^{3}$Department of Chemical and Biological Engineering, Northwestern University, Evanston, IL, USA \\
$^{*}$Correspondence to: \href{mailto:dskoda@ucla.edu}{dskoda@ucla.edu}; \href{mailto:jlopez@northwestern.edu}{jlopez@northwestern.edu}
}
}

\usepackage[resetlabels]{multibib}
\newcites{Supp}{Supplementary References}
\newcites{Main}{References}

\newcommand{\icm}{\mathrm{cm}^{-1}}

\begin{document}

\maketitle
\thispagestyle{firstpagestyle} % Draws the header on the first page

\begin{abstract}
    Interpreting spectroscopy data is a critical bottleneck in automating chemical research and industrial characterization.
    Particularly within infrared (IR) spectroscopy, identifying compounds in complex, liquid-phase chemical mixtures largely relies on expert knowledge, as variable peak assignment, broadening, and shifts hinder data-driven methods.
    Here, we show that an algorithmic approach can  identify components in both simulated and experimental mixture spectra with high accuracy despite nonlinearities in liquid-phase IR data.
    The method is comprehensively benchmarked with a dataset of over 44,000 simulated liquid-phase IR spectra for mixtures and achieves up to 90\% accuracy in identifying molecular components across a dataset of binary and ternary liquid mixtures.
    Our strategy is robust to perturbation of spectra, and its accuracy is capped by near-identical liquid-phase IR spectra that limit the resolution of chemical identification, imposing theoretical limits on achieving perfect accuracy in structure identification.
    Finally, we apply the method to automatically interpret IR spectra in experimental settings, correctly identifying the components of nearly all samples within a blind study.
    This work provides tools and data to advance automated chemical laboratories through algorithmic interpretation of liquid-phase IR spectra of mixtures.
\end{abstract}

\section{Introduction}
Identifying the constituents of molecular liquids is essential for the study and design of chemical formulations across applications, from biomedical and pharmaceutical research to energy materials. \cite{Griffiths2006,Paschoal2017,BarthHaris2009}
In principle, sufficient and high-resolution characterization data can unambiguously determine all components of a liquid mixture and elucidate their intermolecular interactions.
Infrared (IR) spectroscopy is one of the core tools used in chemistry to identify unknown compounds and functional groups in liquid-phase mixtures.\cite{stuart2004infrared, Smith2009}
The technique is fast, nondestructive, and can be paired with complementary measurements such as nuclear magnetic resonance (NMR) and mass spectrometry (MS) to monitor reactions, analyze materials, and identify products in chemical reactions and chemical processes, including under operando conditions.

Many experimental IR datasets for pure gas- and liquid-phase spectra are available in standardized digital formats through sources such as the NIST Chemistry Webbook,\cite{nist_webbook} the NIST Quantitative Infrared Database, \cite{chu1999nist_ir_database} and the Japanese AIST Spectral Database for Organic Compounds (SDBS).\cite{aist_sdbs}
While these datasets provide reference spectra, their limited coverage of chemical space and restricted accessibility constrain their use for large-scale computational analysis and data-driven modeling.
Consequently, compound identification in molecular mixtures using IR spectra relies on expert-driven workflows, including identifying and interpreting spectral signatures using comprehensive functional group tables, performing simplified density functional theory calculations, and conducting pattern-guided searches over a large space of spectra.\cite{Smith2009}
As chemical laboratory automation and liquid-phase handling continue to advance, the limited ability to automate the interpretation of characterization data becomes the main bottleneck in chemical analysis. 

Automation challenges go beyond data availability to the fundamental physics of vibrational spectroscopy.
Spectral peak positions and intensities arise from vibrational frequencies that are sensitive to local thermodynamic conditions, intra- and intermolecular interactions, and mode-mixing phenomena such as hot bands, overtones, and vibrational coupling.\cite{Smith2009}
As a result, structure identification from IR spectra is better posed in the gas phase than in the liquid phase.
Highly reproducible measurements and sharp and distinct peaks characteristic of the gas phase allow near-unambiguous characterization of compounds. 
In contrast, liquid phase spectra result from a broader distribution of molecular geometries that lead to overlapping normal modes and blurred vibrational features.
Liquid mixtures further encode the molecular environment directly in the IR spectrum: they exhibit peak shifts and broadening relative to their gas-phase counterparts, and a mixture spectrum cannot always be modeled well as a simple weighted sum of its constituents. 
Developing data and methods is therefore essential for automating chemical characterization across phases and components.

Historically, chemometric methods based on partial least squares (PLS) have been widely used for predictive modeling of liquid-phase mixtures from IR spectra across pharmaceutical, fuel, and food applications, \cite{granato2018trends,johnson2006fuelquality,roggo2007nirreview,prata2024honey} with numerous small-scale studies demonstrating effective property prediction or classification. \cite{cunha2017biodiesel,sufriadi2023patchouli,wang2019ftirfuel} 
However, their performance depends strongly on spectral preprocessing, reference measurements, and the chemical scope of calibration data, thereby restricting applicability to a very limited, case-by-case basis.\cite{vanmanen2021units, alibrahim2021dcn,nifhuarain2024milk}
Furthermore, few large-scale, chemically diverse, liquid-phase spectroscopic datasets have been evaluated with PLS to test these limitations.
Partial least squares is used to learn statistical covariances between spectra and properties rather than to enforce a physically constrained mixture model.
As a result, while PLS remains useful for liquid-phase spectroscopic analysis, it functions as a pattern recognition-based regression method as opposed to a molecular identification model capable of explicitly representing mixture composition and component identities. 
Similarly, other least squares methods have been used to deconvolute mixture IR spectra, but only for narrow chemical domains.\cite{Haaland:85,Haaland:82,Cahn:88,xiao1989iterative,griffin2014robust,livanos2016deconvolution}

More recently, gas-phase synthetic spectral datasets and machine learning (ML) have been developed to perform structure elucidation from experimental spectra.\cite{alberts2024irstructure, alberts2025benchmarks, lu2024deeplearning, enders2021}
Such tools have led to establishing spectral-to-structure maps in the gas phase and have supported limited structure prediction from pure-component IR spectra. \cite{zipoli2025multimodal,kanakala2024spectratostructure,french2025seq2seq, alberts2024irstructure,alberts2025benchmarks} 
Additionally, data-driven models have been shown to successfully predict the components and concentrations of gas-phase mixtures from spectra with moderate to high accuracy,\cite{jin2024deductive} with only limited demonstrations extending to liquid-phase synthetic and experimental data.\cite{ficarra2025lmmixtures}
Synthetic spectral generation has also advanced through neural network models capable of predicting infrared spectra from molecular structure in the gas phase to support inverse spectral analysis. \cite{abdulal2024nnir, kartha2025,mcGill2021}
Despite these advances, accurate molecular identification from liquid-phase spectra remains insufficient for reliable use in practice.

This context highlights two knowledge gaps. 
On the data side, broader and standardized datasets are needed to understand and evaluate the performance of liquid-phase IR analysis and to enable training newer ML models in the high-data regime.
On the modeling side, the limits of traditional algorithms for deconvolving liquid-phase IR data remain unclear beyond narrow chemical spaces, leaving even baseline performance unestablished.
In this work, we develop a dataset containing over 44,000 simulated, liquid-phase IR spectra to quantify peak shifting in chemical mixtures, and we develop algorithms to automate chemical identification from mixture spectra.
We show that the non-negative least squares (NNLS) algorithm correctly identifies compounds in both gas- and liquid-phase simulated and experimental IR mixture spectra.
Interestingly, the NNLS algorithm does so with high accuracy despite nonlinearities in mixing.
This result remains robust to spectral perturbations, including noise and artificial peak shifts applied to the pure-component spectra.
Component identification reaches 100\% in gas-phase mixtures but is limited in liquid-phase IR, with an optimistic performance of up to 90\% accurate identification.
We show that identification accuracy is bounded not by algorithmic performance, but by near-identical liquid-phase IR spectra that yield degenerate solutions to mixture deconvolution.
By characterizing these degenerate solutions, we indicate a potential theoretical limit to spectral deconvolution that may require additional measurements or information to accomplish component identification in the liquid phase.
We show how our strategy can be useful in practice to deconvolve mixture IR spectra from multiple experimental samples.
In a small-scale blind study where experimental sample identities were withheld during analysis, our algorithm correctly identified the compounds in nearly all experimental samples, demonstrating its applicability in practical settings.
In addition to the dataset and quantitative baselines, we discuss potential limitations to the scalability of characterization and molecular identification in automated laboratories.

\section{Results}

\begin{figure}[htb!]
   \centering
   \includegraphics[width=\textwidth]{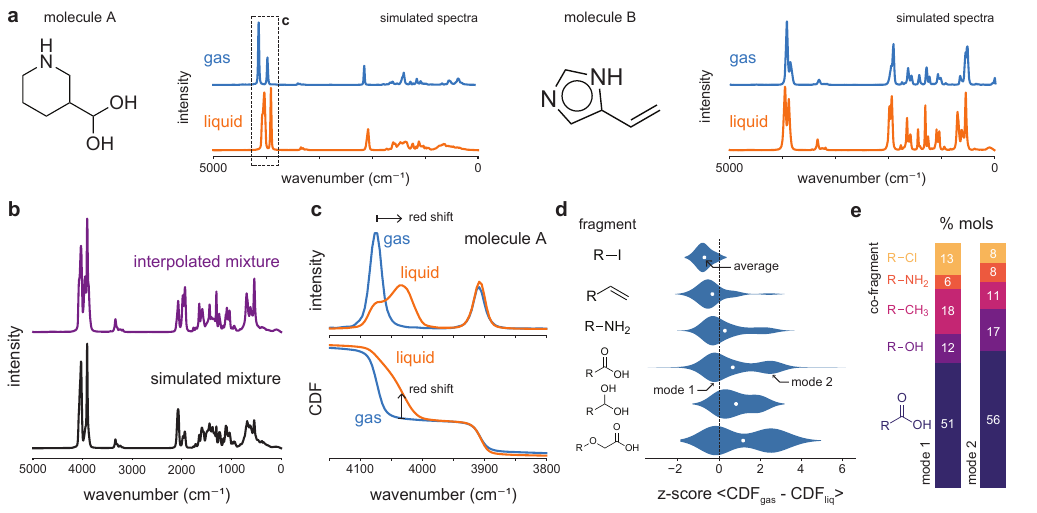}
   \caption{{MD-generated gas- and liquid-phase pure and mixture IR spectra and cumulative intensity difference metric analysis.}
   \textbf{a}, Simulated gas and liquid spectra for 3-(dihydroxymethyl)piperidine (molecule A) and 4(5)-ethylimidazole (molecule B).
   \textbf{b}, Simulated mixture spectrum of a liquid-phase mixture of molecule A and B (bottom) and the equal weight linear combination (average) of the two molecular spectra (top). The linear sum is not equivalent to the true simulated mixture.
   \textbf{c}, Raw and cumulative intensities of molecule A's gas- and liquid-phase spectra between 4150--3800 $\icm$, illustrating the gas-to-liquid peak shift and broadening.
   \textbf{d}, Fragment-driven differences between gas and liquid spectra. Distributions and per-core means of fragment-level z-scores for the cumulative distribution function (CDF) between gas- and liquid-phase spectra. Molecules are decomposed into a Murko-scaffold "core" and their largest remaining fragment. Average CDF values are standardized (z-score) within each core, removing core-specific effects and isolating fragment-dependent contributions.
   The white dots represent the average of each distribution.
   \textbf{e}, Mode-specific relative composition of the most common co-fragments for molecules containing a carboxylic acid fragment (O=CO). Molecules are assigned to one of two modes by fitting a two-component Gaussian to their per-core z-score cumulative intensity differences.
   }
   \label{fig:fig1}
\end{figure}

\subsection{Rationalizing peak shifts in simulated gas- and liquid-phase IR spectra}

To quantify and rationalize the magnitude of peak shifts in simulated IR spectra of gas- and liquid-phase molecules, we created a dataset with 8880 pure gas-phase and 8550 pure liquid-phase spectra using classical simulations (section \ref{simulation-methods}).
Examples of simulated gas- and liquid-phase spectra for 3-(dihydroxymethyl)piperidine (molecule A) and 4(5)-ethylimidazole (molecule B) are shown in Fig. \ref{fig:fig1}a.
When comparing the gas- and liquid-phase spectra across both pure components and mixtures, several features can be observed.
Liquid-phase spectra exhibit spectral broadening relative to the sharp peaks observed in the gas phase in the fingerprint region between 0 and 2000 $\icm$ for both molecules.
For the A + B mixture, the IR spectrum simulated using equal molar amounts of each molecule is not equivalent to the sum of their pure-component liquid-phase spectra, as shown in Fig. \ref{fig:fig1}b.
This shows how intermolecular interactions in the mixture can obscure clear peak assignments of individual components in complex mixtures and give rise to nonlinear mixing behavior characteristic of the liquid phase.

In addition to these mixture effects, liquid-phase spectra also show clear shifts in peak positions relative to their gas-phase counterparts.
Figure \ref{fig:fig1}c depicts how the gas-phase peak around 4100 $\icm$ for molecule A red-shifts when simulated at the liquid phase.
To systematically quantify the magnitude and sign of peak shifts, we normalized each molecule's gas- and liquid-phase spectra to unit area and computed the difference between their cumulative distribution functions (CDFs), where the sign measures the direction of the shift (section \ref{cdf-diff}).
Using the average distance between CDFs, we then analyzed the dataset for trends in peak shifts and broadenings between both spectra. 
To probe the structural dependence of these phase-dependent spectral shifts, each molecule was decomposed into its Murcko scaffold "core" and its largest remaining fragment.
Figure \ref{fig:fig1}d highlights how the average CDF difference captures fragment-dependent peak shifts that are consistent with known phase-dependent intermolecular interaction differences.
Molecules whose largest fragments are hydrocarbons or halogens exhibit smaller average CDF differences between their gas- and liquid-phase spectra, consistent with their weaker intermolecular interactions and lack of hydrogen bonding.
In contrast, fragments such as amines, carboxylic acids, and alcohols show substantial peak shifts and broadening from the gas phase to the liquid phase.

Notably, for some fragments, the distribution of z-scores of the mean CDF differences is bimodal, indicating two subpopulations with different interaction strengths.
Decomposing the molecules in these modes into their full fragment compositions reveals distinct compositional differences between the modes, as illustrated in Fig. \ref{fig:fig1}e. Molecules in mode 1 contain more co-fragments that contribute weakly to gas-liquid spectral differences; for instance, methyl and chlorine fragments co-appear more frequently in mode 1 than in mode 2, explaining their lower average shifts.
Conversely, molecules in mode 2 contain more carboxyl, alcohol, and amine co-fragments, which account for their higher average peak shifts.
This demonstrates that peak shifts in simulated liquid-phase IR spectra can be rationalized according to known intermolecular interaction trends, while the emergence and structure of their distributions are not predicted by chemical intuition alone. 

\begin{figure}[htb!]
   \centering
   \includegraphics[width=\textwidth]{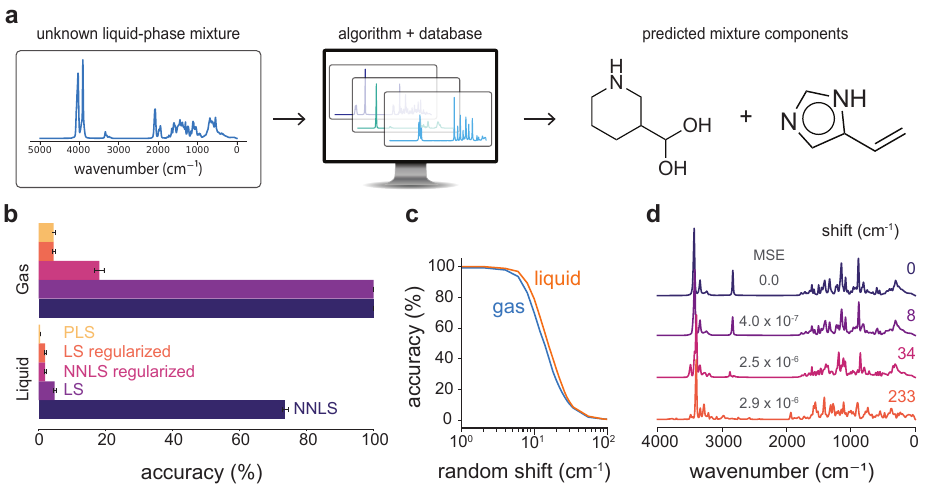}
   \caption{{Identification accuracies of unknown components of two-component liquid-phase mixtures from simulated IR spectra, using MD-generated simulated pure-component IR spectra.}
   \textbf{a}, Workflow to identify unknown mixture components from a liquid-phase mixture spectrum. Given an unknown spectrum and our database of pure-component spectra, an algorithm is used to predict mixture components.
   \textbf{b}, Prediction accuracies using NNLS, LS, and regularized variants to identify two-component mixtures. Gas- and liquid-phase mixtures were predicted using both gas and liquid pure-spectra basis sets, with NNLS achieving the highest liquid-phase accuracy.
   \textbf{c}, Prediction accuracies using NNLS for gas and liquid phase mixtures as a function of spectral peak shifts. 
   \textbf{d}, Examples of spectra with increasing peak shift magnitudes (in $\icm$).
   }
   \label{fig:fig2}
\end{figure}

\subsection{Robustness of deconvolution of simulated IR spectra of mixtures to small shifts}

Given that IR spectra of mixtures often deviate from linear interpolations of pure-component spectra due to liquid-phase peak shifts (Fig. \ref{fig:fig1}), decomposing molecular mixtures into individual components is typically deemed non-trivial.
We therefore tested whether linear algorithmic approaches can succeed in predicting molecular components from an unknown IR spectrum even in the presence of non-linearities in the data.
Figure \ref{fig:fig2}a illustrates the main method tested in this work, where components of each mixture spectrum are predicted using linear algorithms and a ``basis set'' of pure-component spectra.
Using the gas-phase pure-component spectra as the basis set, the components of 30,000 gas-phase and 26,996 liquid-phase mixtures were predicted.
Using the liquid-phase basis set, molecular constituents of 27,657 two-component and 7,985 three-component liquid-phase mixtures were predicted.
Predicted mixture components were obtained by ranking the coefficients by magnitude and selecting the top two (binary) or three (ternary) values, matching the known number of components in each mixture.

Each pure-component basis set included spectra for all molecules present in the mixtures, along with an equal number of spectra from molecules not appearing in any mixture, selected at random.
Each dataset of mixture-basis set combinations was evaluated eight times, using a different random selection of additional molecular spectra in each iteration, shown in Fig. \ref{fig:fig2}b.
Gas-phase mixtures are predicted with the highest accuracy, reaching up to 100\% accuracy with zero standard deviation across prediction runs.
This result reflects the linear additivity of gas-phase spectra: sharp and distinct peaks arising from non-interacting molecules make each spectrum perfectly distinguishable, allowing linear unmixing algorithms to recover the exact components under a linear mixing assumption.
For liquid-phase mixtures, using pure gas-phase spectra as the basis set yields a prediction accuracy of only 15.4\% (Fig. \ref{fig:si_2comp_acc}), indicating that gas-phase spectral signatures differ too substantially from their liquid-phase counterparts to serve as effective reference data for deconvolving condensed-phase mixtures.
This limitation persists even when the gas-phase data are artificially broadened (Fig. \ref{fig:si_2comp_acc}).
When using a basis set of pure liquid-phase spectra that captures liquid-phase spectral effects, NNLS identifies liquid-phase mixture components with 73.6\% accuracy, outperforming all other algorithms.

Given the reasonable success of NNLS in identifying mixture components despite peak shifts arising from gas-to-liquid phase mixing behavior, we quantified the robustness of this algorithm by deliberately introducing random shifts into liquid-phase pure-component spectra used to construct mixtures (section \ref{shift-peaks-method}).
Figure \ref{fig:fig2}c shows that liquid-phase mixtures tolerate slightly larger frequency shifts than gas-phase mixtures.
While any peak shift is expected to reduce identification accuracy, the greater tolerance observed for liquid-phase deconvolution under moderate wavenumber shifts is consistent with broader liquid-phase spectral features, which make them less sensitive to small positional shifts.  
In contrast, sharper gas-phase peaks would be expected to be more vulnerable to small positional perturbations that disrupt features essential for accurate deconvolution.
However, gas- and liquid-phase identification accuracy degrades at similar rates, with accuracy remaining above 80\% for random peak shifts up to 8 $\icm$. 
While both phases show a rapid decline in accuracy beyond approximately 15-20 $\icm$, gas-phase identification accuracy remains consistently lower than that of the liquid phase across all shifts, reflecting greater sensitivity to deviations in peak positions. 
Figure \ref{fig:fig2}d illustrates how small shifts still preserve many of the original spectral features, while larger shifts produce pronounced spectral changes that coincide with the observed loss in identification accuracy.

\begin{figure}[htb!]
   \centering
   \includegraphics[width=0.45\textwidth]{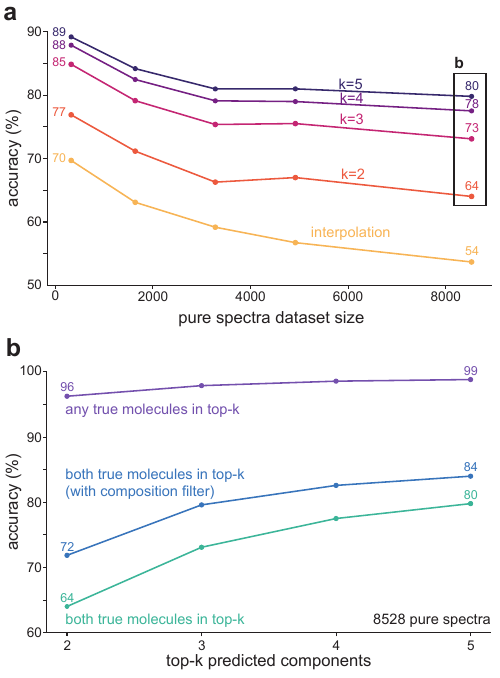}
   \caption{Two-component liquid-phase mixture identification accuracies obtained with NNLS as a function of pure liquid-phase spectra dataset size and prediction criteria.
   \textbf{a} Identification accuracies as a function of pure liquid-phase basis set size. Accuracy is reported for identifying all true components from the largest $k=2$-$5$ NNLS coefficients compared with the interpolation baseline, which selects the top two coefficients from a brute-force convex interpolation over all spectrum pairs. For $k=2$, NNLS achieves higher identification accuracy than the interpolation baseline across all dataset sizes.
   \textbf{b} Identification accuracies as the prediction criterion increases from $k=2–10$ are evaluated by (i) requiring all true components to appear within the top $k$, (ii) requiring at least one (any) true component appears within the top $k$ coefficients, and (iii) applying an atom-count filter that restricts candidate components whose combined atomic compositions match the mixture’s atom count (as would be available from mass spectrometry (MS)).}
   \label{fig:fig3}
\end{figure}

\subsection{Accuracy limits for deconvolution of IR spectra of mixtures with linear methods}
\label{sec:acc-limits}

Considering that NNLS can deconvolve spectra of liquid-phase mixtures with meaningful accuracy (Fig. \ref{fig:fig2}b), we assessed the limits of structure identification with this method.
First, correctly identifying all components of a mixture is the strictest measure of accuracy for an algorithm.
In practice, however, identifying a set of candidate molecules with similar spectral features (and thus structural motifs) can already support indirect peak assignment and interpretation of spectral data.
Second, linear algorithms rely on a spectral database (Fig. \ref{fig:fig2}a), whose size can vary and constrain identification accuracy.
To quantify the limits of accuracy under relaxed criteria and dataset sizes, we evaluated two accuracy metrics: (1) both true molecules are within the top-$k$ molecules predicted by the algorithm; or (2) any of the true mixture components is within the top-$k$ molecules predicted by the algorithm. 
Figure \ref{fig:fig3}a compares how the accuracy of NNLS varies as a function of $k$ and the dataset size under criterion (1).
At the smallest dataset size (i.e., the minimum number of pure components needed to fully identify all mixtures), NNLS correctly predicts 77\% of mixtures when $k=2$, corresponding to exact recovery of the mixture.
Accuracy increases to 89\% when both true molecules are within the top-5 candidates.
When the dataset of pure spectra is increased to the maximum size in this study of 8528 spectra (26 times larger than the minimum dataset size needed), NNLS identifies both true molecules with 64\% accuracy when $k=2$ and with 80\% accuracy when $k=5$.
The algorithm’s performance first plateaus before declining to lower accuracies, indicating the increasing difficulty of identifying the correct components as the candidate space grows. 
On the other hand, the ``brute-force'' interpolation approach, which evaluates all molecular pairs independently (section \ref{sec:deconvolution-method}), resulted in a monotonic decrease in accuracy from 70\% to 54\%.

At the largest dataset size, we examined the NNLS accuracy under a range of evaluation strategies to characterize achievable performance.
Incorporating an atom-type filter during evaluation substantially improved performance at the largest pure dataset size, increasing the top-$2$ accuracy from 64\% to 72\% (Fig. \ref{fig:fig3}b).
Computationally, this filter prioritizes pairs whose elements are known to be present in the mixture, narrowing the pool of candidate molecules (supplemental methods \ref{atom-match-method}).
Experimentally, the presence or absence of atoms that frequently confound NNLS (such as I vs Cl) is often known. 
In cases of true unknown identification, elemental analysis can readily provide the required information to distinguish between two candidate molecules.
To test whether the observed accuracy limit reflects NNLS failing to recover the true components entirely, we assessed whether either of the correct molecules is recovered among top-$k$ candidates predicted by the algorithm.
In this case, Figure \ref{fig:fig3}b shows that the accuracy of the algorithm is nearly perfect, at 99.1\% for $k=5$, demonstrating that NNLS almost always identifies at least one of the true components from the mixture IR spectrum.
Practically, this accuracy is sufficient for dramatic acceleration of interpretation of IR spectra compared to traditional manual functional group analysis. 
The small remaining accuracy gap suggests that further gains may be limited not only by the algorithm, but also by the underlying spectral data.

\begin{figure}[htb!]
   \centering
   \includegraphics[width=\textwidth]{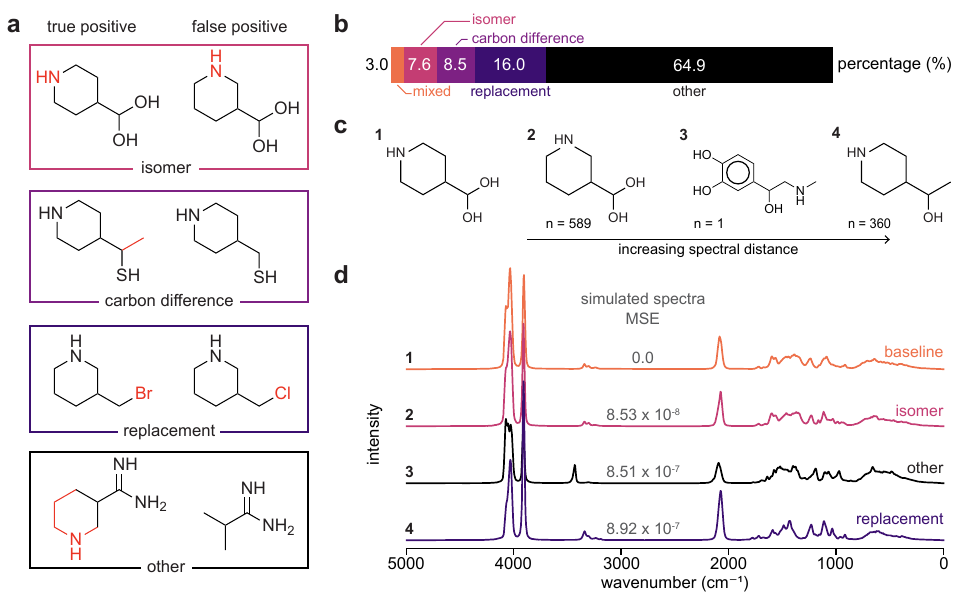}
   \caption{{Misidentification profiles for predicting all components in two-component liquid-phase mixtures using the top $k=2$ NNLS coefficients.}
   \textbf{a}, True vs. falsely predicted components for characteristic misidentification examples: (i) a predicted component differs by the addition or removal of a carbon relative to the true component; (ii) a predicted component is an isomer of the true component; (iii) a predicted component differs by one-atom substitution; and (iv) misidentification not covered by (i)-(iii).
   \textbf{b}, Percentage of two-component mixtures that were misidentified when using the top $k=2$ NNLS coefficients to identify both components, aggregated across pure-component dataset sizes. "Mixed" indicates mixtures where multiple misidentification categories (carbon difference, isomer, substitution) apply to the true-false component pair.
   \textbf{c}, Molecules closest to the true component relative to 3-(dihydroxymethyl)piperidine (molecule A) by spectral distance (MSE) and their corresponding spectra. The spectral similarity among these molecules makes them ambiguous to the NNLS algorithm that minimizes squared error (MSE-equivalent). The variable ``n" indicates the number of times the nearest neighbor molecule was incorrectly predicted instead of the true component, across all evaluated mixtures and pure-spectra dataset sizes.
   }
   \label{fig:fig4}
\end{figure}

\subsection{Characterizing misidentification profiles and spectral ambiguity}
\label{sec:misidentification}

To explain the accuracy limits in Fig. \ref{fig:fig3}b, we analyzed which molecules are often misidentified by NNLS.
We categorized false positive misidentifications according to classes depicted in Fig. \ref{fig:fig4}a.
False positives tend to occur due to (1) structural isomers, i.e., molecules with the same chemical formula but different atomic arrangements; (2) single atom substitutions, where one element is replaced by another; and (3) molecules that differ only in total carbon count where the false positive has one more or one fewer carbon than the true positive. 
Together, these cases explain about 35\% of all false positives predicted by NNLS (Fig. \ref{fig:fig4}b). 
The remaining 65\% of false positives fall into the ``other'' category and involve molecules that share similar molecular characteristics despite being structurally dissimilar.
One such example is 2-methylpropanimidamide and piperidine-3-carboximidamide (Fig. \ref{fig:fig4}a), two carboximidamides that are structurally distinct yet have very similar simulated IR spectra.
The common C=N bond and amine functional groups result in similar major features of the spectra, while the characteristic peaks of the piperidine ring overlap with the broad amine peak (3200 - 3300 $\icm$) and C-H/C-C peaks.
These overlaps result in a mean squared error (MSE) difference of $7.3\times 10^{-7}$ $\icm$ between the true and false positive spectra. 

Figures \ref{fig:fig4}c,d further illustrate this spectral ambiguity by comparing the first three nearest spectral neighbors of a representative molecule (piperidin-4-ylmethanediol), identified using the MSE between liquid-phase IR spectra. 
The resulting MSE values, on the order of $10^{-8}-10^{-7}$, are much smaller compared to the average MSE between two arbitrary molecules, $7.16\times10^{-6}$ $\icm$. 
These small errors illustrate how distinct molecular structures can produce nearly indistinguishable liquid-phase IR spectra, leading NNLS to assign false-positive identifications. 
Figure \ref{fig:fig4}d emphasizes the degree of spectral similarity among these nearest neighbors, particularly the ``isomer'' and ``replacement'' misidentification spectra, where differences are primarily limited to changes in relative peak intensities.
When small structural variations make molecules effectively indistinguishable in liquid-phase IR spectroscopy, misidentification reflects limitations in the discriminative information available from spectra rather than limitations of the prediction algorithm itself. 
This suggests that the accuracy limits reported in Fig. \ref{fig:fig3} arise from intrinsic constraints of linear mixture deconvolution applied to liquid-phase IR spectra.
Thus, misidentification of liquid-phase IR data must be interpreted carefully, even though accuracies approaching 90\% already enable reasonable automation of molecular structure deconvolution.

\begin{figure}[htb!]
   \centering
   \includegraphics[width=0.5\textwidth]{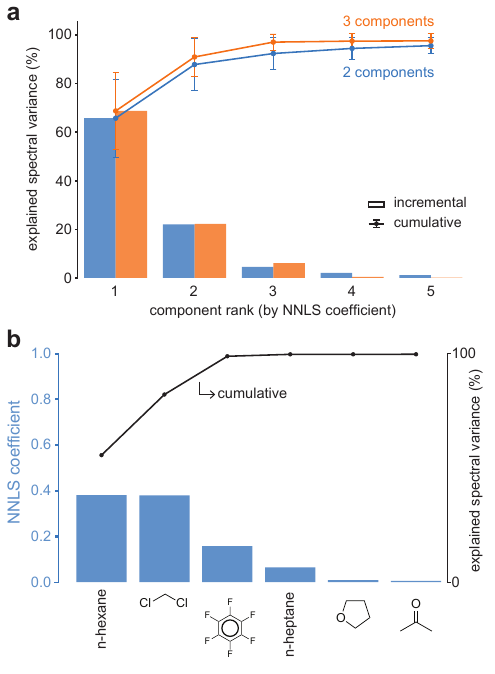}
   \caption{{Fraction of mixture spectra explained by components ranked by decreasing NNLS coefficient for two- and three-component liquid mixtures.}
   \textbf{a}, Average cumulative and incremental percentage of the explained spectrum across all two- and three-component liquid-phase mixtures, ranked in decreasing NNLS coefficient order, as a function of basis set size with associated errors. 
   \textbf{b}, NNLS coefficients for the top six components and the percentage of the spectrum explained for a three-component liquid mixture. The plateau in the explained-spectra curve indicates the likely number of components in the mixture, as additional component spectra weighted by their coefficients do not further contribute to explaining the mixture. 
   }
   \label{fig:fig5}
\end{figure}

\subsection{Interpreting NNLS coefficients to evaluate component contributions and identify n-component mixtures}   

Although the analysis thus far has been performed on binary mixtures, the methods are, in principle, applicable to mixtures containing an arbitrary number of components.
Using 7,985 simulated spectra of three-component liquid-phase mixtures, the NNLS algorithm correctly identified all three components within the top $k=3$ candidates with an accuracy of $73\%$ (Fig. \ref{fig:si_3comp_acc}).
Despite the presence of similar misidentification profiles discussed in Section \ref{sec:misidentification}, the identification accuracy of NNLS for these mixtures remains comparable to the results shown in Fig. \ref{fig:fig3} for two-component mixtures.
This suggests that the algorithm can be applied for automated identification pipelines without further modification.

When the number of components in a mixture is unknown, the coefficients obtained from NNLS deconvolution can be used to infer both the number of components present and their relative contributions to the resulting mixture IR spectrum.
Starting from a spectrum with zero intensity, pure component spectra are added sequentially in decreasing NNLS coefficient order, each weighted by its predicted coefficient.
Then, the cumulative variance of the total spectrum that is explained after each added component hints at the likely number of true components (Fig. \ref{fig:fig5}a).
Once all real components have been included, the explained variance fraction plateaus and additional components yield minimal improvement.
Figure \ref{fig:fig5}b exemplifies this behavior. 
Although at least six molecules have non-zero coefficients, the cumulative explained variance saturates after the third component. 
Additional metrics described in the Supplementary Information (section \ref{component-analysis}) suggest that, within reasonable signal-to-noise thresholds, $n$-component mixtures may be analyzed even when the number $n$ is unknown \textit{a priori}.

\subsection{Structure identification for experimental mixture IR spectra}
\label{sec:experimental-id}
\begin{figure}[htb!]
   \centering
   \includegraphics[width=\textwidth]{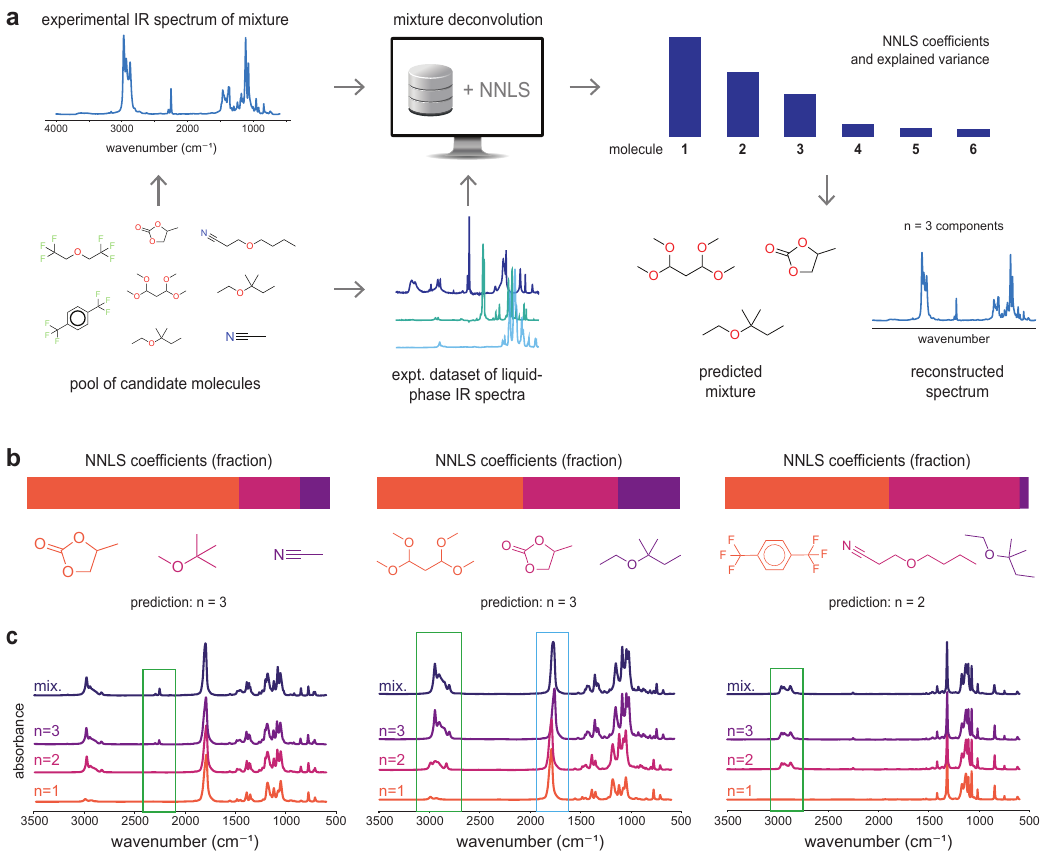}
   \caption{
   Automatic identification of experimental mixtures in a blind study.
   \textbf{a}, Components of liquid-phase mixtures were identified from experimentally observed mixture spectra using a basis set of experimentally measured pure liquid-phase spectra and the approach developed in this work.
   \textbf{b}, Examples of top-3 mixture components ranked by NNLS coefficients for three different mixtures.
   \textbf{c}, Cumulative weighted spectrum reconstructions obtained by sequentially adding the top $n=1,2,3$ NNLS-ranked component spectra, each weighted by its NNLS coefficient. The spectrum labeled as ``mix'' is the measured IR spectrum for the mixture.}
   \label{fig:fig6}
\end{figure}

\begin{figure}[htb!]
   \centering
   \includegraphics[width=\textwidth]{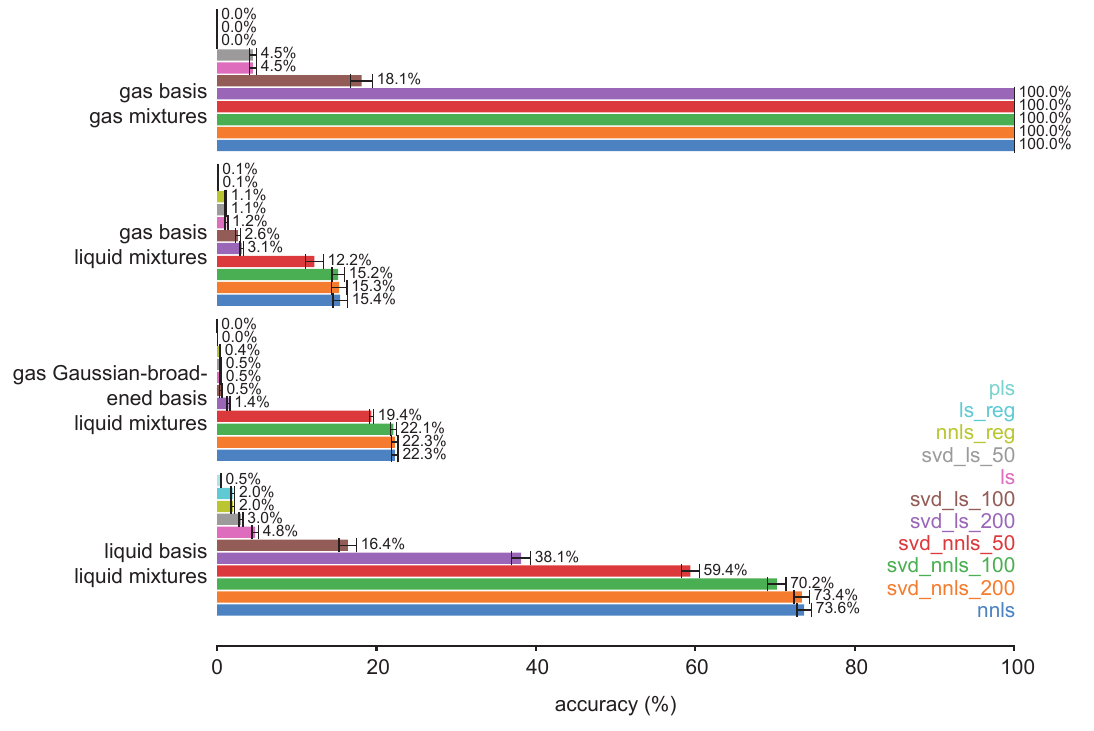}
    \caption{Identification accuracies of unknown components of two-component liquid-phase mixtures from simulated IR spectra, using MD-generated simulated pure-component IR spectra, as described in the main text, Section ~\ref{sec:evaluation1}. Algorithm labels follow the format \textit{alg\_n} where \textit{alg} denotes the coefficient estimation method used to solve $Y$=$CX$, and $n$ (when present) indicates the dimensionality of the singular value decomposition (SVD) latent space onto which $X$ and $Y$ were projected prior to coefficient estimation. Methods without a numeric suffix operate in the original feature space. 
    }
   \label{fig:si_2comp_acc}
\end{figure}

\begin{figure}[htb!]
   \centering
   \includegraphics[width=\textwidth]{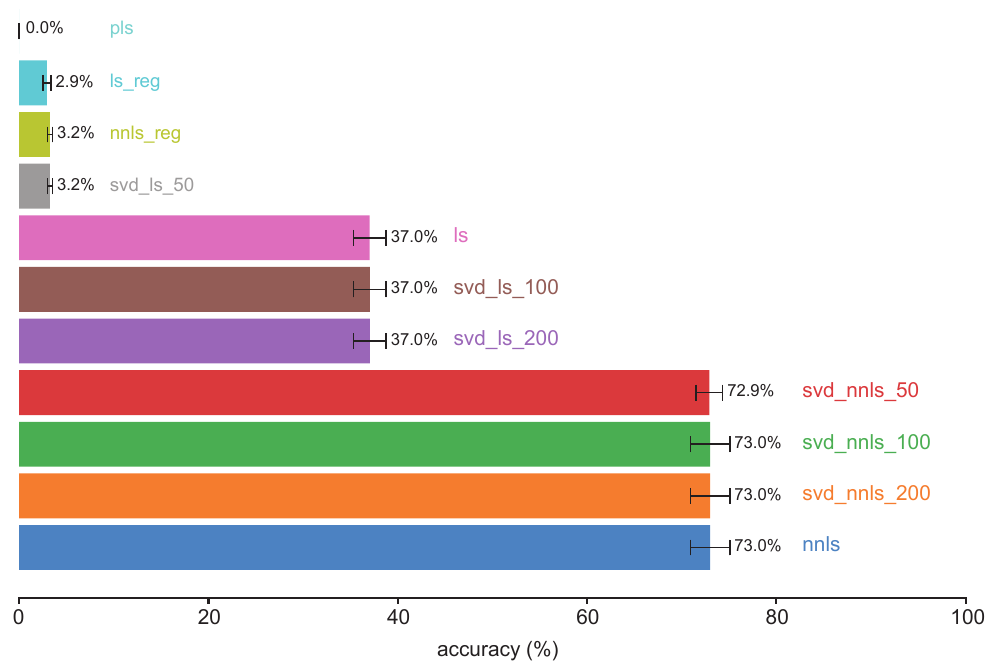}
    \caption{Identification accuracies of unknown components of three-component liquid-phase mixtures from simulated IR spectra, using MD-generated simulated pure-component IR spectra. Algorithm labels follow the format \textit{alg\_n} where \textit{alg} denotes the coefficient estimation method used to solve $Y$=$CX$, and $n$ (when present) indicates the dimensionality of the singular value decomposition (SVD) latent space onto which $X$ and $Y$ were projected prior to coefficient estimation. Methods without a numeric suffix operate in the original feature space. 
    }   
   \label{fig:si_3comp_acc}
\end{figure}

To demonstrate the applicability of the present analysis beyond simulated spectra, we performed a blind study to predict both the identities and number of components in experimentally prepared two- and three-component liquid-phase mixtures.
Our experimental team prepared nine different mixtures and provided IR spectra of pure compounds and mixtures from a pool of common laboratory solvents, as shown in Fig. \ref{fig:fig6}a.
Then, the computational team ranked and predicted the components of the mixtures using the methods discussed in previous sections.
After unblinding the results, we noticed that the NNLS approach accurately identified all true components within the top $k=2$ or $k=3$, effectively deconvolving the real mixture spectra when reference pure-component spectra are available.
Figure \ref{fig:fig6}b illustrates predictions for three experimental mixtures along with the NNLS coefficients used to infer how many components are present in each mixture.
Figure \ref{fig:fig6}c showcases how the reconstructed IR spectra change as components are added sequentially, weighted by their NNLS coefficient.
For example, in the three-component mixtures (left and middle panels in Figs. \ref{fig:fig6}b,c), a peak present in the mixture spectrum appears only after the third component spectrum is added to the cumulative reconstructed spectrum (green box).
On the other hand, in the two-component mixture case (right), characteristic mixture peaks are already reproduced after the first two component spectra are added.
The second mixture in Figs. \ref{fig:fig6}b,c also exhibits a peak shift in the C=O stretch near 1600 $\icm$ between the cumulatively weighted spectrum including all three components and the experimental mixture (blue box), indicating that the third component introduces a shift consistent with the observed experimental mixture spectrum and is accurately captured by the linear algorithm.
Thus, beyond correct identification, analysis of component contributions and coefficient rankings provides a framework for interpreting complex experimental mixtures that enables automated identification of liquid-phase mixtures in laboratory settings.

\section{Discussion}

Interpreting liquid-phase IR spectra often requires identifying nonlinearities that can be explained by intermolecular interactions or chemical arguments, hindering automation of spectroscopic analysis.
Despite the existence of these nonlinearities in spectral data, this work shows that linear decomposition algorithms are sufficient to automatically identify liquid-phase mixtures from IR spectra provided knowledge of the pure-component liquid-phase spectra.
The accuracy of the approach seems to be limited by the degeneracy of IR spectra in the liquid phase rather than nonlinearities in the data.
We established a benchmark for chemical identification in liquid-phase mixtures and highlighted the necessity of liquid-phase spectral data for reliable chemical automation, using a non-negative least squares (NNLS) algorithm and an extensive dataset of simulated spectra.

The presented results demonstrate that the methods can be transferred directly to experimental data, as shown in the identification of candidate components in real experimental mixtures. 
NNLS coefficients provide chemically informative signals, indicating partial spectral similarity to true components and revealing plausible structural motifs that guide interpretation of spectroscopic data (section \ref{component-analysis}).
Importantly, NNLS accurately identifies two- and three-component mixtures and remains robust to peak shifts and to increases in pure-component dataset sizes while preserving chemically interpretable coefficients. 
Poor identification accuracy when using gas-phase pure-component spectra as the ``basis set'' for liquid-phase mixture deconvolution highlights the need to create larger liquid-phase IR spectral datasets.
Existing spectroscopic datasets for liquid-phase IR are either limited in size or are protected by licensing requirements that prevent their use in large-scale identification efforts, while the simulated liquid-phase IR spectra here exhibit systematic deviations from experimental data due to the approximations of the force fields.
Future efforts to curate and share experimental liquid-phase IR spectral data can be invaluable for automation of chemical identification in laboratory or industrial settings.

The MD simulations used to generate our spectral dataset capture liquid-phase spectral behavior, even though the spectra do not exactly reproduce experimental measurements.
The MD-generated IR spectra exhibit self-consistent peak positions, shifts, and broadenings that mirror experimentally observed gas-liquid spectral differences.
These intermolecular interaction-driven differences were quantified using cumulative distribution functions of spectral intensities, confirming that the simulations capture the intermolecular effects that distinguish gas- and liquid-phase spectra.
Importantly, these gas-liquid shifts correlate with known chemical functionality, with molecules that promote stronger intermolecular interactions exhibiting larger spectral shifts. 
Interaction-driven features also persist in simulated mixture spectra, indicating that the simulations capture the anharmonic and nonlinear mixing behavior characteristic of the liquid phase. 

Interpreting NNLS coefficients enables deeper analysis into the deconvolution of mixture IR spectra. 
While accuracy is high in gas-phase identification, misidentification profiles of liquid-phase IR spectra show that prediction failures arise primarily when the IR spectra of true and false-positive candidates are nearly indistinguishable. 
In these cases, misidentifications reflect fundamental limits in discriminative IR information rather than just shortcomings of the linear inversions (section \ref{cdf-analysis}).
The close correspondence between spectral similarity and shared structural motifs of pure components further supports this interpretation.
Furthermore, linear unmixing becomes slower and harder for mixtures composed of highly degenerate spectra, which may reflect challenges in coefficient convergence when weights must be distributed across many nearly indistinguishable candidate spectra.
Even though non-linear methods, including machine learning strategies, could exhibit better performance in component prediction, there will be numerical limits to the accuracy due to degenerate spectra.

Our results show that NNLS deconvolution for liquid-phase mixture identification is ready for use in automated laboratories.
The approach offers interpretable coefficients, robustness to peak shifts, and applicability across both synthetic and experimental data, while providing transparent analysis of spectral reconstruction quality and chemical interpretability (sections \ref{deconvolution-behavior}, \ref{component-analysis}).
The practical value of this interpretable framework is reinforced by further improvements in identification accuracy when incorporating mixture atomic composition information, which is easily available in experimental settings. 
Future work can address the limitations of linear algorithms and dataset sizes to increase the reliability of predictions and further automate the interpretation of nonlinearities in mixture IR spectra. 
Explicitly quantifying spectral degeneracy provides a pathway toward defining the fundamental limits of IR-based mixture identification under any deconvolution algorithm, thereby enabling uncertainty estimates that are particularly valuable in experimental applications.

\section{Conclusion}

This work presented a framework for the automatic identification of components in liquid-phase mixtures using IR spectra and algorithmic approaches that are useful for chemical research and industrial processes.
Automatic identification was performed on a dataset of simulated binary and ternary mixture spectra using a database of simulated pure-component gas- and liquid-phase spectra. 
In particular, we showed that linear deconvolution using the non-negative least squares (NNLS) algorithm enabled component identification from IR spectra of mixtures with up to 90\% accuracy.
The NNLS algorithm yields component coefficients that are interpretable as relative contributions to the mixture spectrum and remain informative even in the cases of false-positive component identifications.
Misidentification profiles were shown to stem from spectral and chemical similarity between candidate compounds, originating from the degeneracy between IR spectra rather than shortcomings of the linear decomposition method.
Finally, this method was shown to be applicable to experimental settings in a blind study on experimentally prepared mixtures, in which the framework successfully identified the mixture components. 
Consequently, this work defines a benchmark for liquid-phase mixture identification from IR spectra and establishes a scalable workflow for automated mixture identification.
Future advances in algorithmic methods, spectral modeling methods, and the availability and expansion of spectral databases can continue to improve automated interpretation of complex chemical mixtures.

\clearpage
\section{Methods}

\subsection{Simulation methods}
\label{simulation-methods}
\textbf{Molecular simulations:}
Gas- and liquid-phase pure component spectra along with liquid-phase two- and three- component mixture spectra were generated by molecular dynamics (MD) simulations using OpenMM\cite{eastman2023openmm} and a custom-made code.
The potential energy surface was represented using OpenFF's Sage (v. 2.0.0) force field, as implemented in the OpenForceField codebase.\cite{boothroyd2023development,wagner2021openforcefield,mobley2018escaping}
Simulations were initialized by instantiating an OpenFF Molecule object using the SMILES string, and creating a conformer in the gas phase using OpenFF's in-built functions.
All simulations were performed at a constant temperature of 300 K, enforced by a Langevin thermostat with a coupling constant of 1 ps$^{-1}$, and a time step of 2 fs.
Although the high time step leads to peak shifts corresponding to C-H, O-H, and N-H bonds, the shifts are systematic across the dataset and do not influence the identification methods.
For liquid simulations, simulation boxes were initialized with a constant density of 0.5 g/cm$^3$ containing 100 molecules using PACKMOL.\cite{martinez2009packmol}
All mixture simulations were performed in the liquid phase with equimolar ratios, thus with 50 molecules of each species inside the simulation box.
A Monte Carlo algorithm implemented in OpenMM was used as a barostat in the equilibration stage of the liquid phase simulations to simulate the effect of constant pressure, with a frequency of 200 fs$^{-1}$.
Throughout all simulations, equilibration was performed for 0.5 ns for every system.
Afterwards, production simulations were performed for 0.5 ns in the NVT ensemble.
In liquid-phase simulations, the equilibrium density of the box was taken as the average density of the last 100,000 steps of the equilibration trajectory.
These simulations showed good convergence with respect to density and IR spectra, and thus were adopted as default parameters for this study.
However, approximations related to the force fields prevent direct comparison with experimental data across this study.

\noindent\textbf{Calculation of the IR spectra: }
To compute the IR spectra consistently for both gas- and liquid-phase molecules, the net molecular dipole of the simulation box was computed at each timestep.
Then, the total system dipole was recorded in a file for each time step.
The IR spectrum was obtained by following the implementation from Braun,\cite{Braun2016Zenodo,alberts2024irstructure} taking the Fourier transform of the autocorrelation function of this dipole over the trajectory to compute the frequency-dependent absorbance:
\begin{equation}\label{eq:ir-fourier}
\tilde{C}(\nu) = \int_{0}^{\infty} C(t)\cos(2\pi \nu t)\,dt,
\end{equation}

\begin{equation}
C(t) = \frac{\left\langle \vec{M}(0)\cdot\vec{M}(t)\right\rangle}{\left\langle \vec{M}(0)\cdot\vec{M}(0)\right\rangle}.
\end{equation}

where $\vec{M}(t)$ is the total dipole moment of the system at time $t$, $\nu$ is the frequency, and $\langle \cdot \rangle$ denotes an ensemble average over the trajectory.
To account for coupling with an electromagnetic field and thermal weighting of vibrational modes, the spectrum is modified with correction factors:

\begin{equation}
F(\nu) = \nu \left( 1 - e^{-\frac{\hbar \nu}{k_B T}} \right)
\end{equation}

\begin{equation}
Q(\nu) = \frac{\nu}{1 - e^{-\frac{\hbar \nu}{k_B T}}}
\end{equation}
where $\hbar$ is the reduced Planck constant, $k_B$ is the Boltzmann constant, $T$ is the temperature, and $F(\nu)$ and  $Q(\nu)$ are the field description and quantum correction factors, respectively, used to adjust the classical spectrum to a quantum corrected spectrum:
\begin{equation}
S(\nu) = \tilde{C}(\nu)\, F(\nu)
\end{equation}
\begin{equation}
S_{\mathrm{qm}}(\nu) = S(\nu)\, Q(\nu)
\end{equation}

The final expression for the IR spectrum is thus given by:
\begin{equation}
S_{\mathrm{qm}}(\nu) = \tilde{C}(\nu)\, \nu^2
\end{equation}

Gas-phase spectra were generated using a single molecule per simulation box with molecules considered sufficiently far apart so that intermolecular dipoles are effectively uncorrelated.
Under these conditions, mixtures of gas-phase spectra are assumed to have vanishing correlations between the dipoles of different molecules.
Thus, the IR spectrum of a gas-phase mixture is simply the mole-fraction weighted linear combination of the pure-component spectra.

\textbf{Automated calculation workflow:}
Automation of the IR spectra calculations was implemented using the mkite software.\cite{schwalbekoda2023mkite}
The software suite allows for the generation of jobs with an arbitrary number of inputs in a matrix-like fashion, thus allowing the calculation of two- and three-component mixtures within a single database.
Data management was performed using a new plugin, developed for this work at \url{https://github.com/mkite-group/mkite-infrared} (available at publication time).

Every spectrum is represented by a one-dimensional vector of 1250 wavenumber indices, corresponding to a uniform 4 $\icm$ grid spanning 0 to 5000 $\icm$, with every value being the intensity at each wavenumber.
The raw MD-derived intensities were interpolated onto this common grid, Gaussian smoothed to reduce noise, and negative intensities where clipped to zero.
Each spectrum was normalized by its total integrated intensity to produce a probability-density-like spectrum vector.

\textbf{Dataset of molecules:}
%gas total : 8800
%liq total: 8550
% 2comp liq total: 27870
%3comp liq total: 7985
%total = 44,405
%2 comp liq for deconvolution (components have pure liq): 27657
%3 comp liq for deconvolution - 7985
The dataset of about 8880 molecules used for calculations of gas- and liquid-phase spectra was created by combining the AqSolDB \cite{sorkun_aqsoldb_2019} (originally with 9,982 compounds) with two other custom-made datasets.
One custom dataset was about 50 molecules that were readily accessible to the experimental team.
The other dataset was around 380 molecules generated by combinatorially matching one molecular ``core'' scaffold with one fragment.
To generate the combinatorial core-fragment library, a curated set of structurally diverse fragments, defined as monovalent R-group substituents, was combined with a collection of reactive core scaffolds. 
The core scaffolds comprised related structural backbones that differed in the position of the fragment attachment site, enabling single-point R-group modification at distinct locations.
All possible combinations of the defined scaffold core and R-group fragments were enumerated to produce a dataset of closely related yet structurally distinct molecules.
The SMILES of the resulting molecules were canolicolized and subjected to chemical sanitization to ensure data consistency, valence correctness, and structural validity. 

After removing structures for which MD simulations failed to converge, we obtained 8,880 pure gas-phase and 8,550 pure liquid-phase simulations.
Mixture molecules were selected using two approaches.
From the AqSolDB dataset, binary and ternary mixtures were constructed by randomly selecting pairs or triplets of compounds. 
From the core-fragment dataset, mixtures were generated by enumerating all possible pairwise combinations. 

\subsection{Experimental IR spectra}
\label{sec:main:experimental-ir-spectra}
Three two-component and six three-component experimentally prepared liquid-phase mixtures were evaluated with all algorithms using 143 liquid-phase spectra as the basis set.
The basis set contained twenty pure components that were true components of the mixtures, measured using Fourier transform infrared spectroscopy with attenuated total reflectance (FTIR-ATR).
The ATR apparatus (Pike instrument), using diamond as the ATR prism, was connected to the IR spectrometer (Nicolet iS50, Thermo Scientific) with a deuterated triglycine sulfate (DTGS) detector.
The incident IR light was fixed at 45$^{\circ}$.
The IR background and sample spectra (4000 to 400 $\icm$ range) were collected with a spectral resolution of 4 $\icm$ and averaged over 128 scans.
ATR correction was performed in OMNIC software (Thermo Scientific).
Ethylene carbonate (EC, Gotion), dimethyl carbonate (DMC, Sigma-Aldrich), diethyl carbonate (DEC, Sigma-Aldrich), ethyl methyl carbonate (EMC, Gotion), fluoroethylene carbonate (FEC, Gotion), propylene carbonate (PC, Sigma-Aldrich), acetonitrile (Thermo Scientific), tert-butyl methyl ether (Sigma-Aldrich), tert-amyl ethyl ether (Sigma-Aldrich),
1,4-bis (trifluoromethyl) benzene (TCI America), 1,1,3,3-tetramethoxypropane (Sigma-Aldrich), 3-butoxypropionitrile (Sigma-Aldrich), bis(2,2,2-trifluoroethyl)ether (Synquest Labs, Inc.), 2,2,2-trifluoroethyl acetate (Synquest Labs, Inc.), hexafluorobenzene (Sigma-Aldrich), 1,3-dioxolane (DOL, Sigma-Aldrich), ethylenediamine (TCI America), ethyl methyl sulfone (EMS, TCI America), 1,2-dimethoxyethane (DME, Sigma-Aldrich), and 1,1,2,2-tetrafluoroethyl 2,2,3,3-tetrafluoropropylether (TTE, TCI America) were used as received.
An additional 123 pure-component spectra were obtained from the Open Specy IR spectral library,\cite{Cowger2021} to augment the basis set and increase prediction difficulty.

During evaluation, the true number of mixture components was hidden.
The cumulative fractional spectral variance, mean squared error, and algorithm coefficients (described in sections \ref{sec:evaluating-deconvolution} and \ref{sec:spectral_metrics}) were used to predict mixture components and infer the number of components.

\subsection{Deconvolution algorithms for IR spectra data}
\label{sec:deconvolution-method}
In this work, we treated single-component IR spectral data as the ``basis set'' of a multi-component mixture IR spectrum.
Least squares (LS), non-negative least squares (NNLS), and their regularized variants were tested to estimate component contributions in simulated mixture spectra.

The LS method minimizes the cost function

\begin{equation}\label{eq:algo-ls}
    \min_{\mathbf{C}} \; \| \mathbf{Y} - \mathbf{X}\mathbf{C} \|_2^2 + \lambda \norm{\mathbf{C}}^2
\end{equation}

\noindent whereas the NNLS method minimizes the same cost function while imposing a non-negative restriction on the coefficients,

\begin{equation}\label{eq:algo-nnls}
    \min_{\mathbf{C} \ge 0} \; \| \mathbf{Y} - \mathbf{X}\mathbf{C} \|_2^2 + \lambda \norm{\mathbf{C}}^2
\end{equation}

\noindent In Eqs. \eqref{eq:algo-ls} and \eqref{eq:algo-nnls}, $\mathbf{Y}$ denotes the set of simulated mixture spectra and $\mathbf{X}$ denotes the set of simulated pure component spectra. 
The coefficient matrix $\mathbf{C}$ is defined such that every row corresponds to a mixture and each column corresponds to a component in the basis set.
From this, every value in a mixture row of $\mathbf{C}$ specifies the coefficient of a particular component in that mixture. 

In the non-regularized formulations, the regularization coefficient $\lambda$ is set to zero.

In addition to the algorithms above, a ``brute-force'' interpolation method was also implemented as a baseline for two-component mixtures.
It exhaustively evaluates all pairs of single-component spectra and solves for the least-squares optimal convex mixing coefficients.
The interpolation method serves as a reference for evaluating component coefficient accuracy under the same linear mixing assumption as in the gas phase.

\subsection{Evaluating identification accuracy}
\label{sec:evaluating-deconvolution}

The coefficients corresponding to the single-component spectra in the basis set, as given by the linear algorithms, were used to identify the components of a mixture from its spectrum. 
Identification accuracy was evaluated using two approaches.
In both cases, the coefficients obtained were ranked in descending order of their absolute values, and identification accuracy was evaluated by determining whether the top $k$ coefficients corresponded to the spectra of the true components in the mixture.

\subsubsection{Exact Top-$k$ component identification}
\label{sec:evaluation1}
The top-2 and top-3 coefficients were used for binary and ternary mixtures, respectively. 
In this method, the pure-component basis set included spectra for all molecules present in the mixtures, as well as an equal number of spectra from molecules not appearing in any of the mixtures. 
These spectra were selected at random.
Identification accuracy was evaluated eight times, each using a different random selection of additional molecular spectra.

\subsubsection{Criterion-Based top-$k$ identification}
\label{sec:evaluation2}
Identification accuracy was also evaluated for finding the correct mixture components from the top-$k$ coefficients evaluated by (1) requiring all true components to appear within the top $k$, (2) requiring at least one (any) true component to appear within the top-$k$, and (3) applying an atom-count filter that restricts candidate components whose combined atomic compositions match the mixtures atom composition (described in supplementary section \ref{atom-match-method}).
Using these criteria, identification accuracy was evaluated as a function of pure-component basis set size. 
Only the components contained in all the mixtures were part of the smallest basis set size. 
For larger pure-component dataset sizes, additional pure component spectra were added to the basis set, ensuring that each subsequently large basis set size contained all the pure-component spectra in the smaller basis sets. 

When computing coefficients using varying basis set sizes, NNLS failed to converge for the two largest basis set sizes.
Using a basis set size of 4,920 pure components, NNLS failed for 12 mixtures, corresponding to 0.043\% of the mixture dataset.
Using a basis set size of 8,528, 22 mixtures failed to converge, corresponding to 0.080\% of the mixture dataset. 
Mixtures for which NNLS did not converge were considered unidentifiable and counted as incorrect predictions when accuracy metrics were computed. 
These failures were attributable to the NNLS solver reaching its iteration limit under increased basis collinearity at larger basis sizes.

\subsection{Spectral analysis metrics}
\label{sec:spectral_metrics}
Three metrics were used to quantify a predicted component's contribution to a mixture spectrum: the average cumulative distribution function difference, fractional spectral variance, and mean squared error (MSE).

\subsubsection{Cumulative distribution function difference}
\label{cdf-diff}
The difference between the cumulative distribution functions (CDFs) of two spectra was used to quantify relative spectral differences. 
At a given wavenumber, a larger cumulative intensity difference indicates red shifting of one spectrum relative to the other. 
Cumulative intensity curves can also be used to identify raw intensity differences, reflected by vertical differences between the curves, and relative spectral broadening, indicated by their horizontal differences.
The average CDF difference, defined as the mean difference between two cumulative spectra across all wavenumbers, was used as a quantitative metric for comparing two spectra.

\subsubsection{Fractional spectral variance}

The fractional spectral variance metric measures how well a mixture spectrum is reconstructed as components are added in decreasing order of their NNLS coefficient values.
Starting from a zero-intensity spectrum, pure component spectra were sequentially added in descending order of their NNLS coefficient rankings.
At each addition step, the selected component spectrum was scaled by its predicted coefficient and cumulatively summed to form a partial reconstruction of the mixture spectrum.
After each addition, the partial reconstruction $\hat{y}$ was compared to the true mixture $y$, and a cumulative explained variance metric was computed:

\begin{equation}\label{eq:r2}
R^{2} = 1 - \frac{\lVert y - \hat{y} \rVert^{2}}{\lVert y \rVert^{2}}
\end{equation}

This value measures the fraction of variance in the true spectrum explained by the partial reconstruction.
 
The same procedure was applied using MSE and the average CDF difference to assess each component's contribution to reconstruction quality.
The incremental contribution of each component was evaluated using these three metrics: the stepwise increase in $R^2$, the decrease in MSE, and the decrease in average CDF difference as the mixture spectrum was assembled from zero intensity.

A complementary analysis was performed by removing each component from the basis set, recomputing the NNLS coefficient vector for the same mixture spectrum, and calculating the resulting $R^2$, MSE, and average CDF difference for the predicted mixture spectrum relative to the true spectrum. 
This method quantifies the degradation in fit to the true spectrum when a component is unavailable, indicating the importance of that component's spectral features in explaining the mixture spectrum. 
Component contributions were quantified by the drop in $R^2$ and the corresponding increases in MSE and average CDF difference when that component was removed.

\section*{Data Availability}

The datasets generated in this work will be released upon publication of the manuscript.

\section*{Code Availability}

The code to reproduce the results/plots from this work will be released at publication time.

\section*{Acknowledgements}

Financial support for this publication results from Scialog grant \#SA-AUT-2024-024a from Research Corporation for Science Advancement and Frederick Gardner Cottrell Foundation.
Y.J.U.M. acknowledges supplementary funding from the Amazon AI Ph.D. Fellowship at UCLA Samueli.
This work used computational and storage services associated with the Hoffman2 Shared Cluster provided by UCLA Office of Advanced Research Computing’s Research Technology Group, as well as Delta CPU and Delta GPU at NCSA through allocation MAT240040 from the Advanced Cyberinfrastructure Coordination Ecosystem: Services \& Support (ACCESS) program, which is supported by National Science Foundation grants \#2138259, \#2138286, \#2138307, \#2137603, and \#2138296.
The work performed at the Reactor Engineering and Catalyst Testing (REACT) Core Facility of the Northwestern University Center for Catalysis and Surface Science (CCSS) was supported by a grant from the DOE (DE-SC0001329).  Partial support for instrumentation in REACT is also provided by Northwestern's MRSEC program (NSF DMR-2308691).
This work made use of the Keck-II facility (RRID: SCR\_026360) of Northwestern University’s NUANCE Center, which has received support from the IIN and Northwestern's MRSEC program (NSF DMR-2308691).
The authors thank Gabe Gomes, Jose Regio, and Jiawei Guo for useful discussions.

\section*{Conflicts of Interest}

The authors have no conflicts to disclose.

\section*{Author Contributions}

%Y, T, J, D
\noindent\textbf{Y.J.U.M.:} Methodology; Software; Validation; Formal Analysis; Investigation; Data Curation; Writing -- Original Draft; Writing -- Review \& Editing; Visualization.
\noindent\textbf{T.N.:} Methodology; Validation; Investigation; Writing -- Review \& Editing.
\noindent\textbf{J.L.:} Conceptualization; Methodology; Formal Analysis; Investigation; Resources; Writing -- Review \& Editing; Supervision; Project Administration; Funding Acquisition.
\noindent\textbf{D.S.-K.:} Conceptualization; Methodology; Software; Formal Analysis; Investigation; Resources; Data Curation; Writing -- Original Draft; Writing -- Review \& Editing; Visualization; Supervision; Project Administration; Funding Acquisition.

\clearpage
\beginsupplement

% -----------------------
% Supporting Information
% -----------------------
\beginsuppinfo
\customlabel{sec:sinfo}{Supplementary Information}

\section{Supplementary Text}
\customlabel{sec:stext}{Supplementary Text}

\subsection{Gas-Phase spectra broadening and identification accuracy}
Both gas- and liquid-phase pure component spectra were used as the basis set to deconvolve liquid-phase mixture spectra to identify components. 
When using gas-phase component spectra, the identification accuracy for two component mixtures was found to be 15.4\%, compared to the 73.6\% identification accuracy when using liquid-phase pure component spectra as the basis set, as shown in Fig.~\ref{fig:si_2comp_acc}. 
This result indicates that gas-phase information alone is insufficient for identifying liquid-phase mixtures.
Peaks in each gas-phase spectrum were broadened with a Gaussian kernel to approximate liquid-phase spectral shapes, which increased identification accuracy to 22.3\%. 
This minimal accuracy improvement demonstrates that simple peak broadening cannot compensate for the interaction-driven spectral differences between gas-phase and liquid-phase spectra. 
Gaussian convolution does not significantly improve identification accuracy because it only widens gas-phase peaks.
Without introducing frequency shifts, mode coupling, or intensity differences that characterize liquid-phase spectra, spectral shapes produced from broadening match neither gas-phase spectra nor true interaction-modified liquid-phase features, resulting in poor performance.

\subsection{Identification accuracy across prediction criteria and basis set size}

Under the strictest evaluation criterion, which requires identification of both components from the top-2 NNLS coefficients, expanding the pure-component basis to include spectra not present in the mixtures reduces component identification accuracy (Fig.~\ref{fig:fig3}a). 
The inclusion of additional pure-component spectra expands the solution space of the NNLS decomposition.
This change reduces the relative coefficient dominance of true components among the top $k$ coefficients by introducing more candidate spectra whose spectral features partially reproduce the mixture spectral features. 
Expanding the pure-component spectral basis decreases spectral error because the larger set of candidate spectra gives the NNLS algorithm more flexibility to approximate the mixture by distributing coefficient weight across the additional spectra (Fig.~\ref{fig:si_metric_dataset_size}).
This allows coefficients associated with the true components to decrease in magnitude while permitting them to remain among the top-2.
A false spectrum coefficient would overtake a true one only when the spectral shape is sufficiently close and it reduces the reconstruction error more. 
When the true and false molecules have nearly indistinguishable spectra, the algorithm cannot reduce the squared error by assigning more weight to a true molecule than to a nearly identical false one, leading to an incorrect identification. 
This behavior explains the gradual, not sharp decline in identification accuracy when using NNLS as more pure-component spectra are included in the basis set (Fig.~\ref{fig:fig3}).  
In contrast, the interpolation method selects the single pair of spectra that yields the lowest residual, so any marginally better matching false spectrum fully replaces a true one, causing accuracy to drop more rapidly as the number of available false pure spectra increases.
Evaluating identification accuracy under a less restrictive criterion where both true components are only required to be found from the top-$5$ coefficients increases identification accuracy from 64\% to 80\% with the largest basis set size. 
Non-true mixture components within the top-$k$ correspond to molecules with partial similarity to true components, revealing plausible structural motifs or subcomponents that can guide interpretation and illustrate that NNLS provides meaningful coefficients.
Furthermore, in settings where the number of mixture components is unknown, considering a larger $k$ could be useful for narrowing candidates to a manageable set of chemically related structures.

\subsection{Deconvolution behavior of regularized and non-regularized linear algorithms}
\label{deconvolution-behavior}
The coefficient distributions from regularized and non-regularized algorithms illustrated in Figure ~\ref{fig:si_alg_coeff_errors} show that regularization shifts coefficient values toward smaller magnitudes. 
For both LS and NNLS, regularization compresses the coefficient distribution toward zero, suppressing the magnitude of true component contributions.
In contrast, the non-regularized solutions exhibit broader coefficient distributions.  
For non-regularized NNLS in particular, a clear separation emerges between very small coefficients and a smaller subset of larger coefficients, as shown in Figure \ref{fig:si_alg_coeff_errors}c.
This bimodal distribution structure indicates that non-regularized NNLS distinguishes components whose spectra contribute strongly to the mixture spectrum from those that contribute minimally. 
Regularization suppresses this separation by shrinking larger coefficients and consequently reducing the model's ability to differentiate dominant contributors. 
The consequence of this regularized deconvolution behavior is significantly lower identification accuracy. 

The spectral MSE distributions show that non-regularized solutions achieve substantially lower spectral reconstruction error than their regularized counterparts for both LS and NNLS (Fig.~\ref{fig:si_spectral_errors}). 
This is consistent with the coefficient behavior: when the coefficients are not forced toward zero, sufficient coefficient weight can be allocated to the set of dominant pure-component basis spectra, allowing the mixture spectrum to be reconstructed more accurately. 
By contrast, regularization limits the model's ability to represent components that share many spectral features with the mixture, which leads to increased spectral error.

The coefficient MSE values relative to the ideal molar ratios used to generate the simulated mixture spectra further support this interpretation. In the MD simulations, each mixture was constructed from equal molar amounts. 
Thus, for a two-component mixture, an ideal mixture model would assign coefficients of 0.5 to the two true components, and zero valued coefficients to all other components.
However, the recovery of exactly 0.5/0.5 coefficients is not expected from a linear deconvolution model because the liquid-phase spectra arise from intermolecular interactions and nonlinear mixing effects whose spectra are not linear combinations of the pure-component spectra. 

Within the linear approximation, non-regularized NNLS exhibits the smallest median and mean coefficient error relative to the ideal 0.5 reference value, indicating that the true components' coefficients are closest to the underlying molar composition (Table~\ref{table:coeff}). 
Conversely, non-regularized LS shows larger deviations from ideal mixing.
Least squares can assign negative coefficients to certain basis spectra to reduce the mixture spectrum reconstruction error.
However, these negative coefficients are not physically meaningful, even when the magnitude is used to evaluate deviation from ideal molar compositions.
As a result, LS produces coefficients with larger ideal coefficient MSE than NNLS, suggesting that enforcing non-negativity constrains the resulting coefficient solution toward physically meaningful mixing weights.
The MSE between the ideal and NNLS coefficient values increases over basis set size because candidate molecules whose spectral features can additionally represent the mixture spectrum are allocated coefficient weight, thus reducing the true component coefficient values further from the ideal value of 0.5. 

Table ~\ref{table:nonzero} shows that the deconvolution algorithms produce less sparse (more non-zero) coefficients as the pure-component dataset size increases. 
Notably, NNLS produces sparser coefficient vectors than LS and both of the regularized algorithms, further indicating that using NNLS results in the true component coefficient values deviating less from the ideal coefficient values as more component spectra that can approximate the mixture spectrum are added. 

\begin{table}[ht]
\centering
\caption{Mean squared error summary statistics of the coefficients obtained by solving regularized and non-regularized algorithms versus the ideal coefficients (0.5 for each of the true components).}
\label{table:coeff}
\begin{tabular}{lrlllr}
\toprule
algorithm & basis set size & median & mean & std & n mixtures \\
\midrule
LS & 328 & $4.39\times10^{-2}$ & $5.23\times10^{-2}$ & $4.01\times10^{-2}$ & 27657 \\
LS REG & 328 & $2.47\times10^{-1}$ & $2.47\times10^{-1}$ & $1.05\times10^{-3}$ & 27657 \\
NNLS & 328 & $4.60\times10^{-2}$ & $5.73\times10^{-2}$ & $4.43\times10^{-2}$ & 27657 \\
NNLS REG & 328 & $2.47\times10^{-1}$ & $2.47\times10^{-1}$ & $1.05\times10^{-3}$ & 27657 \\
LS & 1640 & $1.44\times10^{-1}$ & $1.50\times10^{-1}$ & $2.94\times10^{-1}$ & 27657 \\
LS REG & 1640 & $2.48\times10^{-1}$ & $2.48\times10^{-1}$ & $8.98\times10^{-4}$ & 27657 \\
NNLS & 1640 & $5.68\times10^{-2}$ & $6.75\times10^{-2}$ & $4.79\times10^{-2}$ & 27657 \\
NNLS REG & 1640 & $2.48\times10^{-1}$ & $2.48\times10^{-1}$ & $9.02\times10^{-4}$ & 27657 \\
LS & 3280 & $1.82\times10^{-1}$ & $1.77\times10^{-1}$ & $4.55\times10^{-2}$ & 27657 \\
LS REG & 3280 & $2.48\times10^{-1}$ & $2.48\times10^{-1}$ & $8.40\times10^{-4}$ & 27657 \\
NNLS & 3280 & $6.33\times10^{-2}$ & $7.34\times10^{-2}$ & $5.04\times10^{-2}$ & 27657 \\
NNLS REG & 3280 & $2.48\times10^{-1}$ & $2.48\times10^{-1}$ & $8.51\times10^{-4}$ & 27657 \\
NNLS & 4920 & $6.55\times10^{-2}$ & $7.47\times10^{-2}$ & $4.98\times10^{-2}$ & 27645 \\
NNLS & 8528 & $7.24\times10^{-2}$ & $8.06\times10^{-2}$ & $5.14\times10^{-2}$ & 27635 \\
\bottomrule
\end{tabular}
\end{table}

\begin{table}[ht]
\centering
\caption{Average number of non-zero coefficients per mixture for each algorithm and basis set size, where coefficients are considered non-zero if greater than $10^{-6}$.}
\label{table:nonzero}
\begin{tabular}{lrr}
\toprule
algorithm & basis set size & mean non-zero count \\
\midrule
LS & 328 & 166 \\
LS & 1640 & 824 \\
LS & 3280 & 1645 \\
LS REG & 328 & 324 \\
LS REG & 1640 & 1432 \\
LS REG & 3280 & 2491 \\
NNLS & 328 & 19 \\
NNLS & 1640 & 28 \\
NNLS & 3280 & 30 \\
NNLS & 4920 & 33 \\
NNLS & 8528 & 44 \\
NNLS REG & 328 & 324 \\
NNLS REG & 1640 & 1388 \\
NNLS REG & 3280 & 2222 \\
\bottomrule
\end{tabular}
\end{table}

\subsection{Sequential addition and removal metrics for component contribution analysis}
\label{component-analysis}
Quantifying spectral metrics (as described in~\ref{sec:spectral_metrics}) in the presence or absence of specific components allows analyzing multi-component mixtures in terms of relative contributions to predict the component count.
Cases in which a component shows a large NNLS coefficient but a small or even negative incremental contribution indicate that its spectral features are redundant with those of other components. 
Sequential-removal metrics make this redundancy explicit by removing a component and recomputing coefficients to reveal whether other spectra can substitute for its features. 
If the reconstruction improves upon the removal, the removed component was not uniquely informative and could have had broad or overlapping features that can be captured by other spectra that simultaneously capture more of the mixture spectrum. 
Conversely, if the reconstruction worsens, no other spectra can compensate for the features it describes in the mixture spectrum, indicating that it provides unique information about the mixture. 
This coefficient redistribution behavior also offers a way to reason about molecules missing from the basis set that may be present in the mixture. 
When a dominant but incorrect predicted component is removed and the mixture reconstruction improves, the components assigned larger coefficients in the re-solved model approximate the structure of the true but absent molecule more closely than the removed one does, providing structural guidance about unknown mixture components. 

\subsection{Spectral similarity patterns in false component predictions}
\label{cdf-analysis}
Among the misidentified mixtures, the cumulative intensity differences clarify how the NNLS and interpolation component predictions differ in the spectral characteristics that lead to misclassifications.
For each mixture, average cumulative distribution function (CDF) differences were computed between the two true component spectra (true-true pairs), between true and falsely predicted components (false-true pairs), and between two falsely predicted components (false-false pairs), shown in Figure~\ref{fig:si_pure_mix_pairs}a.
This analysis was done for all mixtures misidentified across all basis-set sizes.

For both the false-true and false-false pair types, the NNLS distributions are shifted toward smaller spectral differences compared to interpolation.
Both NNLS and the interpolation method produce false-true pairs that are more spectrally similar than random pairs and are also more similar than the true-true pairs.
Further, the true-true spectral differences are comparable to or larger than the random pairings.
These two results suggest that misclassifications do not arise because the true components in a mixture are inherently too similar but rather due to candidate components being closer in spectral space than typical true pairs.

Importantly, NNLS false-true pairs have smaller average and median CDF differences than the interpolation false-true pairs, indicating that when NNLS selects a false component, it is more spectrally similar to the true component than a false component selected by the interpolation method is. 
The same pattern for false-false pairs is seen as well: NNLS false-false pairs are spectrally closer than false-false pairs from interpolation (and random pairings).
NNLS failures tend to select two false components that are close to one another in spectral space, whereas the interpolation method's two false components resemble random pairings.
This behavior is consistent with the NNLS algorithm that optimizes continuous coefficients over the entire pure-component basis set.
Near-neighbor spectra can receive partial coefficient weighting, resulting in the spectrum of a falsely identified molecule lying close to the correct molecule's spectrum in spectral space. 
Conversely, interpolation evaluates discrete candidate pairs, which leads to incorrect selections that are more spectrally distinct. 

On the mixture level, reconstructed spectra for the misidentified mixtures were generated using the coefficients obtained from the NNLS and interpolation methods. 
Mixtures reconstructed with NNLS coefficients are more spectrally similar to the true mixture spectra than those produced using interpolation coefficients (Fig. ~\ref{fig:si_pure_mix_pairs} b). 
This result is consistent with the pure-component pair analysis.
By design, NNLS retains flexibility to approximate the true mixture spectrum even when the identified components are incorrect because the algorithm optimizes continuous coefficients over the full basis set. 
In contrast, the interpolation method is restricted to discrete candidate pairs and is thus limited in reproducing the mixture spectrum, resulting in larger spectral deviations when misidentifications occur.

\section{Supplementary Methods}
\label{sec:si:methods}

These Supplementary Methods describe in more detail the additional results and calculations in the Supplementary Information.

\subsection{Gaussian broadening of gas-phase spectral peaks}
Gaussian broadening was applied to simulated gas-phase spectra by convolving the spectra with a normalized Gaussian kernel. The kernel was constructed over a window of fixed size, centered at zero, and its width was controlled by the standard deviation parameter, which determines the spread of the Gaussian. 
The Gaussian kernel standard deviation was 20 $\icm$ (corresponding to $\sigma = 5$ grid points on a spectral grid with 4 $\icm$ spacing), yielding a full width at half maximum (FWHM) of 47.1$\icm$.
The kernel was normalized to ensure spectral intensity values were preserved after the convolution was applied, and the output spectrum retained the same length as the original input spectrum. 
This procedure smooths sharp spectral features by redistributing intensity locally across neighboring wavenumbers.
It was used to approximate the peak broadening that is observed between gas-phase and liquid-phase spectra. 

\subsection{Shifting peaks method}
\label{shift-peaks-method}
The peak shifting analysis reported in Fig.~\ref{fig:fig2}c was performed using mixture spectra constructed as linearly weighted sums of pure-component spectra.
Each pure spectrum was modified using a local peak-shifting procedure. 
Randomly selected spectral windows were shifted by amounts sampled from a normal distribution, resulting in local displacements in peak positions while preserving the overall spectral structure.
A smooth blending procedure was applied when inserting the shifted windows to avoid discontinuities in the spectrum, and the spectrum was rescaled to preserve total intensity after shifting. 
The same shifts were applied consistently to each molecule's pure spectrum to preserve its spectral identity.
The NNLS algorithm was then applied to obtain coefficients for the linearly combined mixtures using the shifted pure spectra basis set.  

\subsection{Atom match filtering method}
\label{atom-match-method}
Atom count information for molecules in each mixture was used to filter implausible pure-component predictions obtained using NNLS coefficients (Fig. \ref{fig:fig3}b).
For a given mixture, the total number of each atom type was enumerated from the true mixture composition.
In practice, this elemental composition information could be obtained using mass spectroscopy without knowledge of the specific components present in the mixture.

Assuming an $n$-component mixture (here, $n=2$), all pure-component combinations whose summed atom counts exactly matched the mixture atom counts were identified. 
The set of molecules appearing in any valid combination was taken as the plausible candidate pool for predicting that mixture's components.
The NNLS algorithm was used to obtain coefficients for all pure components, as described in section~\ref{sec:deconvolution-method}.
The coefficients were then restricted to those corresponding to molecules in the plausible candidate pool, and the top $k$ candidates ranked by absolute coefficient magnitude were used to evaluate identification accuracy as described in section~\ref{sec:evaluation2}.

\subsection{Fragment-dependent contributions to gas-liquid spectral differences and mode decomposition}

Each molecule was decomposed into its Murko scaffold ``core'' and the largest remaining fragment. 
Within each core, the average CDF difference value of all molecules sharing that core was standardized (z-scored). 
This per-core standardization removes core-specific effects, so the resulting z-scores reflect fragment-dependent contributions to the gas-liquid spectral difference.
The composition of co-fragments in the two modes for molecules with a carboxylic acid as the largest fragment was analyzed as an example in this work. 
Molecules with this fragment were partitioned into two modes by fitting a two-component Gaussian mixture to their per-core z-scored cumulative intensity differences (the molecule's average CDF difference minus the core-specific mean, divided by the core-specific standard deviation). 
Co-fragment occurrences were counted across molecules in the two modes, and the most common co-fragments were used to calculate relative compositions in each mode.

\clearpage
\section{Supplementary Figures}

\begin{figure}[htb!]
   \centering
   \includegraphics[width=\textwidth]{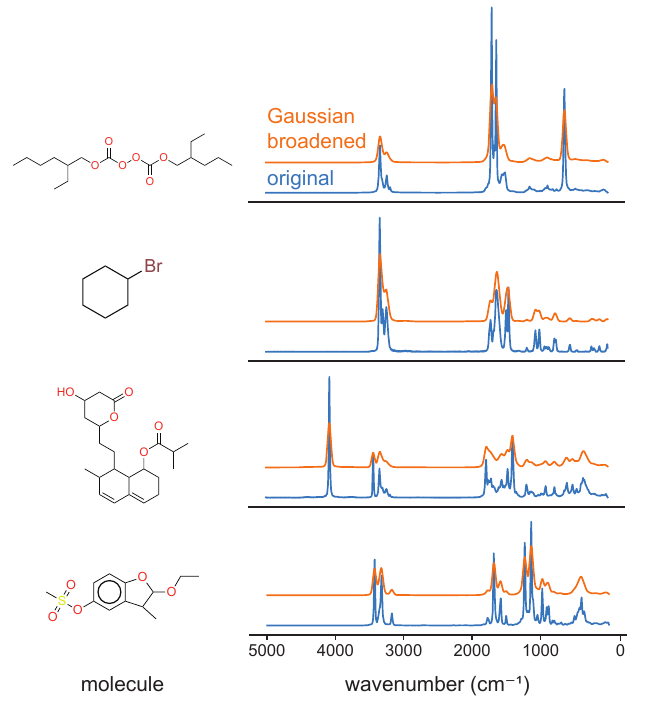}
   \caption{Examples of simulated gas-phase spectra (blue) with their corresponding Gaussian-broadened spectra (orange), as described in the Supplementary Methods ~\ref{sec:si:methods}.
   }
   \label{fig:si_fig3}
\end{figure}

\begin{figure}[hbt!]
    \centering
   \includegraphics[width=\textwidth]{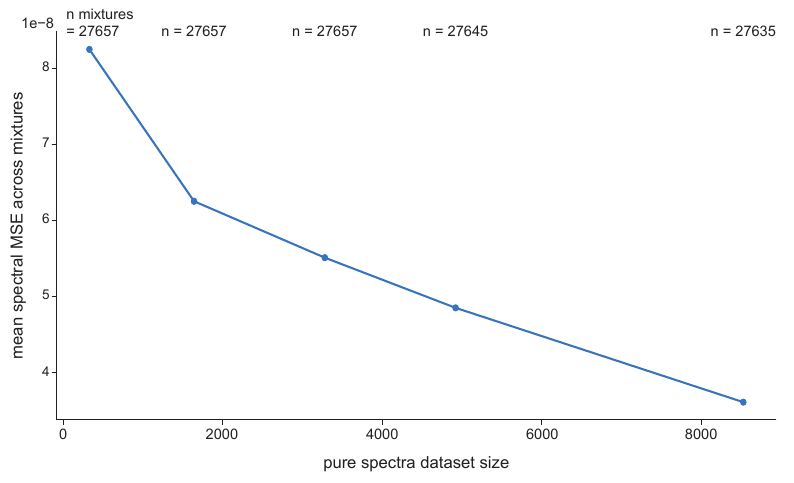}
    \caption{Mean spectral mean squared error (MSE) and mean average cumulative distribution function (CDF) difference across all mixtures for varying pure-component basis set sizes. The number of mixtures, $n$, successfully deconvolved for each basis set size is indicated.
    }
    \label{fig:si_metric_dataset_size}
\end{figure}

\begin{figure}[hbt!]
    \centering
   \includegraphics[width=0.9\textwidth]{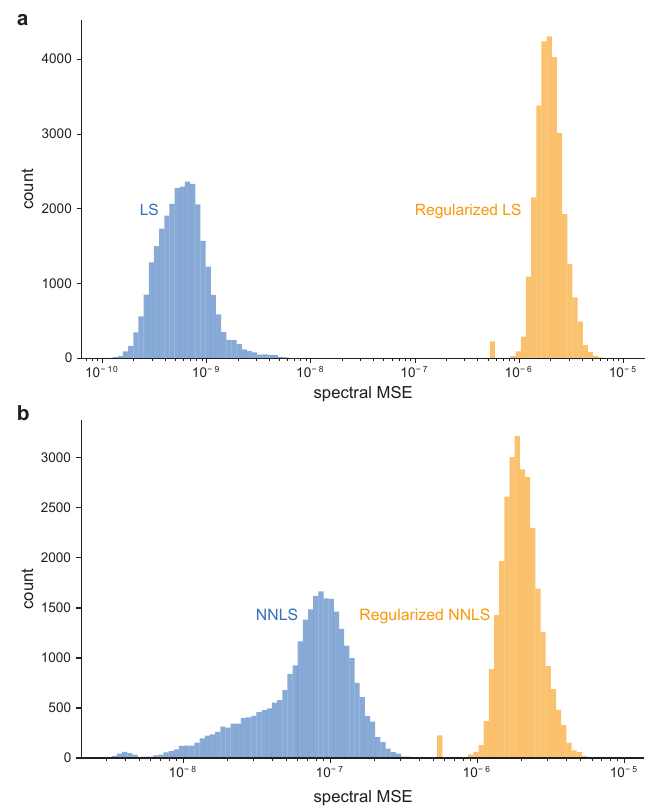}
    \caption{Distributions of the mean squared errors (MSEs) between the true mixture spectrum and the spectrum reconstructed by each indicated algorithm, for all mixtures that were successfully deconvolved in this study across all basis set sizes.
    \textbf{a}, Spectral MSE distributions for the LS and Regularized LS methods.
   \textbf{b}, Spectral MSE distributions for the NNLS and Regularized NNLS methods.
    }
    \label{fig:si_spectral_errors}
\end{figure}

\begin{figure}[hbt!]
    \centering
   \includegraphics[width=0.6\textwidth]{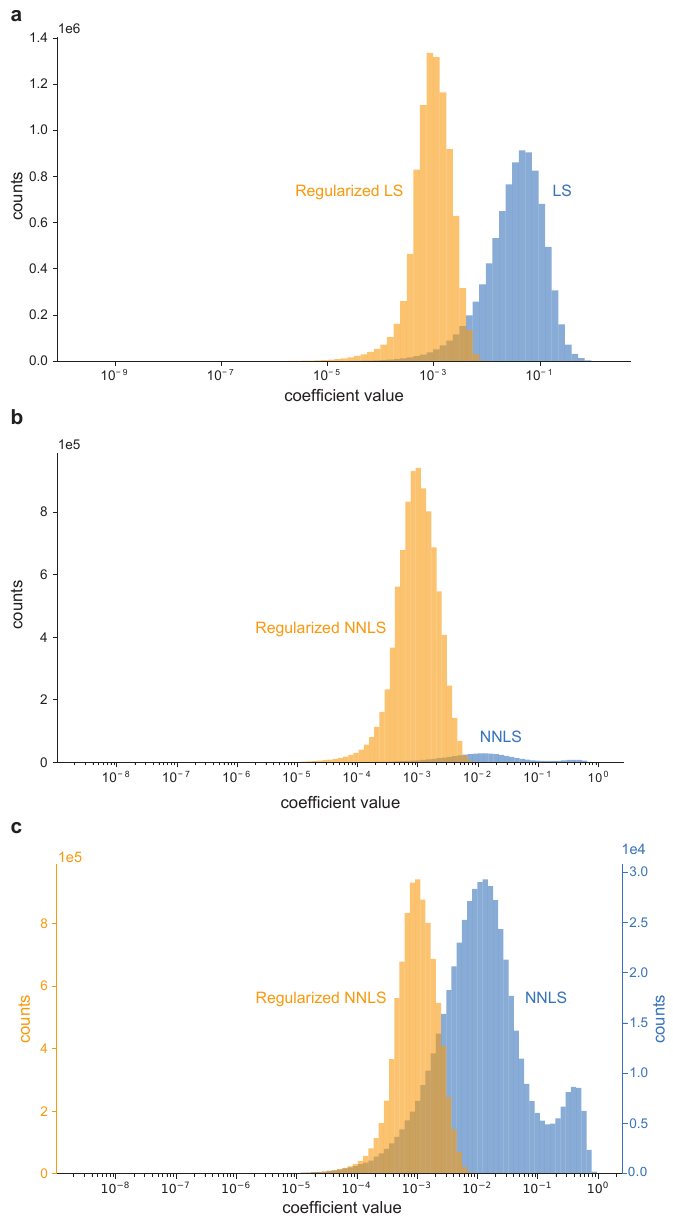}
    \caption{Distributions of the non-zero coefficient values obtained by each indicated algorithm for all mixtures successfully deconvolved in this study across all basis set sizes.
    \textbf{a}, Coefficient distributions for the LS and Regularized LS methods.
    \textbf{b}, Coefficient distributions for the NNLS and Regularized NNLS methods.
    \textbf{c}, NNLS and Regularized NNLS coefficient distributions using separate y-axes for each method to better visualize the shape of the NNLS coefficient distribution. 
    }
    \label{fig:si_alg_coeff_errors}
\end{figure}

\begin{figure}[hbt!]
    \centering
   \includegraphics[width=0.75\textwidth]{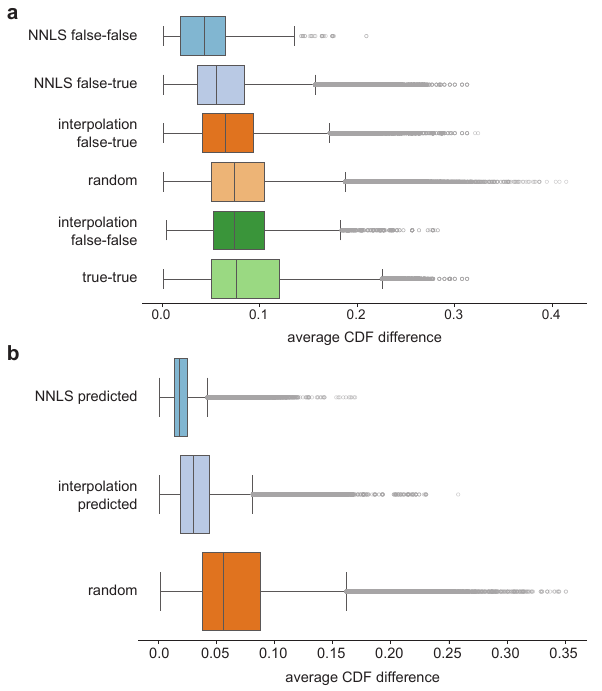}
    \caption{Distributions of spectral differences for misidentified two-component mixtures across all basis-set sizes using the NNLS and interpolation methods.
    \textbf{a}, Average cumulative distribution function (CDF) differences between pure-component spectra. ``True-true'' denotes the two correct components in a mixture; ``false-true'' denotes pairs consisting of one true and one falsely predicted component; and ``false-false'' denotes pairs of two falsely predicted components. Random pairings (200,000 pairs) are shown for reference. NNLS false selections are shifted toward smaller spectral differences relative to interpolation, indicating that NNLS misidentifications involve components that are more spectrally similar to the true molecules. 
    \textbf{b}, Average CDF differences between reconstructed and true mixture spectra using the NNLS and interpolation methods. Mixtures reconstructed by NNLS are more spectrally similar to the ground truth than those reconstructed by interpolation, while random mixtures exhibit larger spectral differences (shown for reference).
    }
    \label{fig:si_pure_mix_pairs}
\end{figure}

\begin{figure}[hbt!]
    \centering
   \includegraphics[width=\textwidth]{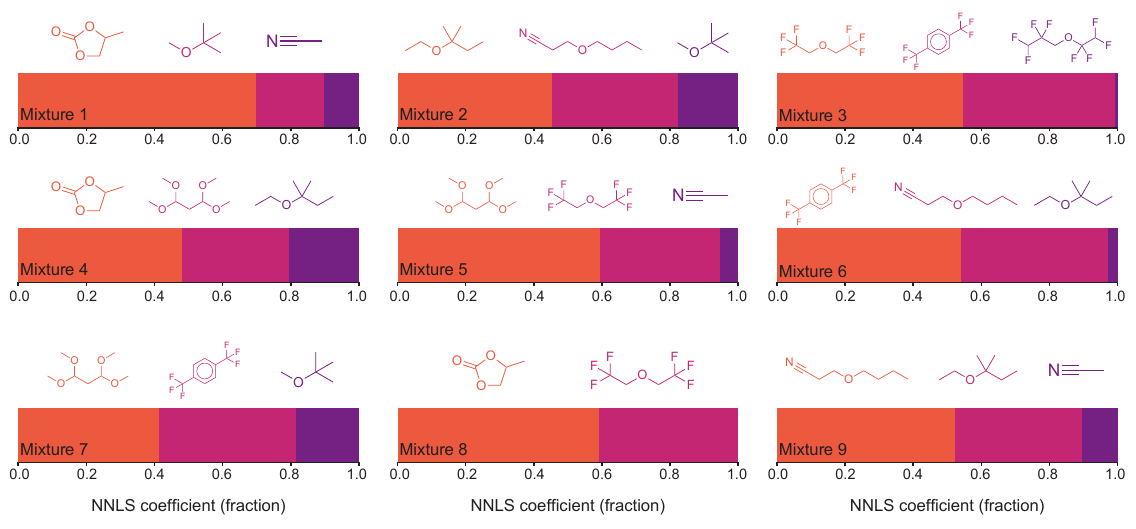}
    \caption{Top-2 and top-3 mixture components ranked by NNLS coefficient for all experimental mixtures, as described in the main text sections ~\ref{sec:experimental-id} and ~\ref{sec:evaluation1}.
    }
    \label{fig:si_exp}
\end{figure}

\begin{figure}[hbt!]
    \centering
   \includegraphics[width=\textwidth]{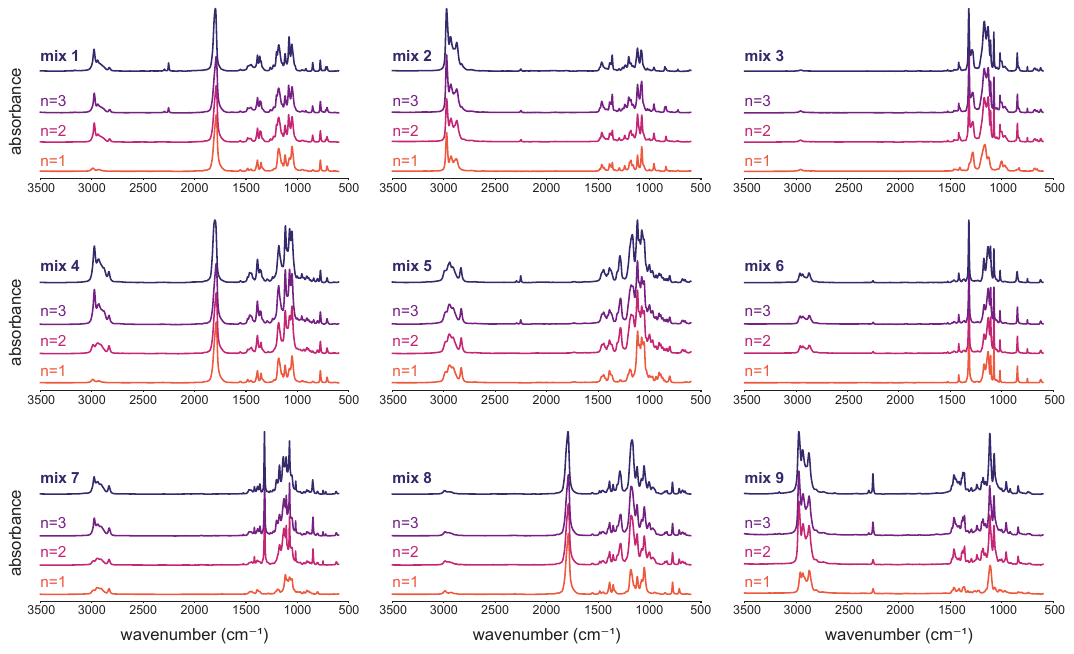}
    \caption{Cumulative weighted spectrum reconstructions for all experimental mixtures, obtained by sequentially adding the top $n=1-3$ NNLS-ranked component spectra, each weighted by its NNLS coefficient.
    }
    \label{fig:si_exp_spec}
\end{figure}

\end{document}